\documentclass[a4paper,11pt]{article}
\usepackage{pos}
\usepackage{lineno}
\usepackage{slashed}
\usepackage{subcaption}
\usepackage{mathtools}
\usepackage{commath}
\usepackage{multirow}
\usepackage{hyperref}
\usepackage{cleveref}
\usepackage{booktabs}
\usepackage{nicefrac} 
\usepackage{comment}
\usepackage{graphicx}
\usepackage{float}
\usepackage{enumitem}
\usepackage{physics}

\title{Electroweak Symmetry Restoration in Extended Higgs Sectors via Domain Walls}

\author[a]{Mohamed Younes Sassi}
\author*[a,b]{Gudrid Moortgat-Pick}

\affiliation[a]{II. Institut für Theoretische Physik,
University of Hamburg,\\Luruper Chaussee 149, 22761 Hamburg, Germany}

\affiliation[b]{Deutsches Elektronen-Synchrotron DESY, Notkestr. 85, 22607 Hamburg, Germany}

\emailAdd{mohamed.younes.sassi@desy.de}
\emailAdd{gudrid.moortgat-pick@desy.de}

\abstract{Domain walls are a type of topological defects that can arise in the
early universe after the spontaneous breaking of a discrete symmetry. This occurs in several beyond Standard Model theories with an extended Higgs sector such as the Next-to-Two-Higgs-Doublet model
(N2HDM). In this talk, I will discuss the domain wall solution related
to the singlet scalar of the N2HDM and demonstrate the possibility of electroweak symmetry restoration (EWSR) in the vicinity of the domain wall. Such symmetry restoration can have profound implications on the early universe cosmology as the sphaleron rate inside the domain wall would, in principle, be unsuppressed compared with the rate outside the wall.}

\FullConference{42nd International Conference on High Energy Physics (ICHEP2024)\\
18-24 July 2024\\
Prague, Czech Republic\\}

\begin{document}

\begin{flushright}
DESY-24-143
\end{flushright}

\maketitle

\section{Introduction}
The matter-antimatter asymmetry of the universe is one of the most important problems in particle physics that cannot be explained by the standard model. Electroweak baryogenesis \cite{Bodeker:2020ghk,Cline:2018fuq}, relying on bubbles of the broken vacuum that are  generated by a first-order phase transition at the electroweak epoch is a well-known mechanism to generate the excess of matter in the early universe. However, this mechanism suffers from stringent experimental constraints on the possible CP-violation needed to satisfy the second Sakharov condition for baryogenesis \cite{ACME:2018yjb}. In this talk, we propose to use the domain walls generated by the real singlet scalar of the N2HDM in order to restore the EW symmetry in a region around the wall. This will lead to a separation of those regions where the weak sphalerons are active (i.e. inside the wall) and exponentially suppressed (i.e. outside the wall). We also show the possibility of generating CP-violating vacua localized on the outer edge of the wall. This will lead to a chiral asymmetry in the fermionic current injected inside the wall. As a consequence, the sphalerons, active inside the wall, generate an excess of baryons over antibaryons. 

\section{Electroweak Symmetry Restoration in the N2HDM}
\begin{figure}[H]
     \centering
     \begin{subfigure}[b]{0.32\textwidth}
         \centering
         \includegraphics[width=\textwidth]{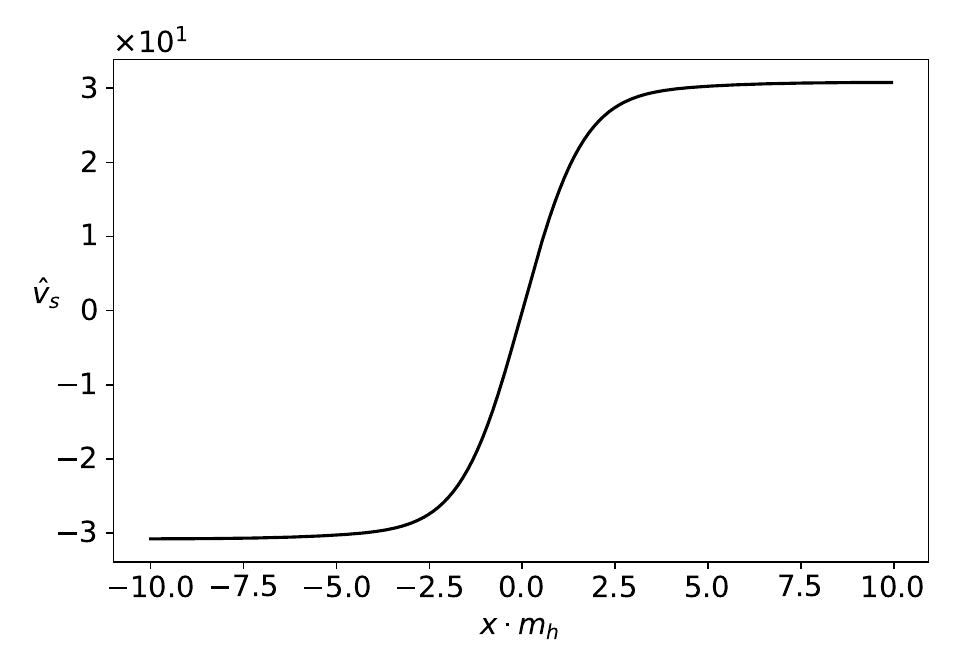}
        \subcaption{} \label{subfig:vs}
     \end{subfigure}     
     \begin{subfigure}[b]{0.32\textwidth}
         \centering
         \includegraphics[width=\textwidth]{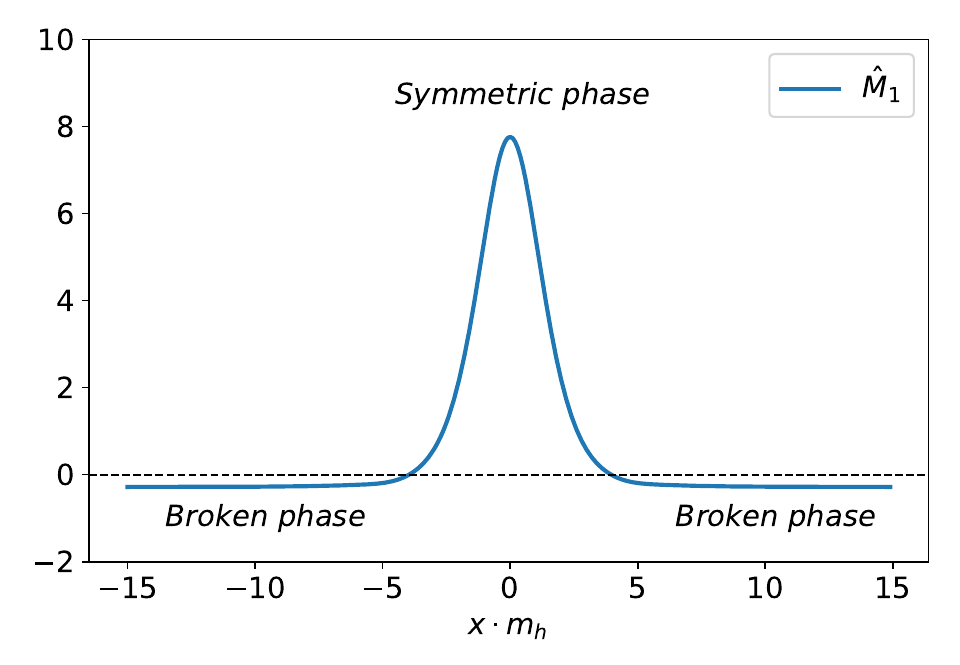}
        \subcaption{} \label{subfig:effmass}
     \end{subfigure}
     \begin{subfigure}[b]{0.32\textwidth}
         \centering
         \includegraphics[width=\textwidth]{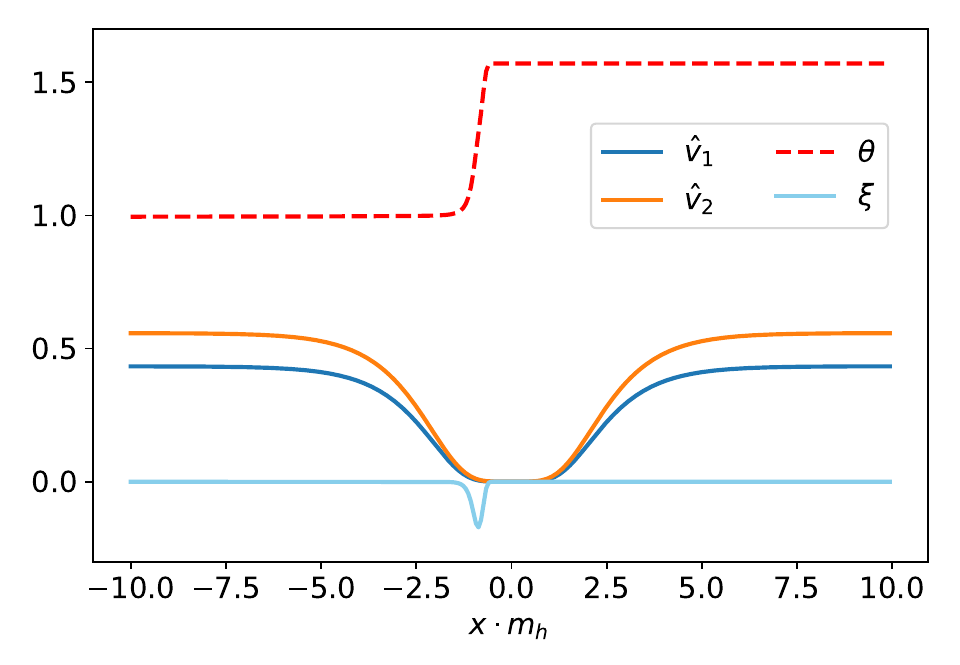}
       \subcaption{}  \label{subfig:sol}
        \end{subfigure}
\caption{(a) Normalized profile of $\hat{v}_s(x) = v_s(x)/v_{sm}$. (b) Normalized effective mass $\hat{M}^2_1(x) = M^2_1(x)/m^2_h$ (with $m_h = 125.09 \text{ GeV}$) as a function of x. (c) Vacuum field profile of the Higgs doublets in the background of the singlet domain wall solution $v_s(x)$.} 
\label{fig:plot}
\end{figure}

The scalar potential considered in this work is given by:
\begin{align}
  \notag & V_{\text{N2HDM}} =  m^2_{11}\abs{\Phi_1}^2 + m^2_{22}\abs{\Phi_2}^2 + (m^2_{12}\Phi^\dagger_1\Phi_2 + h.c.) + \frac{\lambda_1}{2}\abs{\Phi_1}^4 + \frac{\lambda_2}{2}\abs{\Phi_2}^4   + \lambda_3\abs{\Phi_1}^2\abs{\Phi_2}^2 \\ & + \lambda_4\bigl(\Phi_1^{\dagger} \Phi_2\bigr)\bigl(\Phi_2^{\dagger} \Phi_1\bigr)  
    +(\frac{\lambda_5}{2}\bigl(\Phi_1^{\dagger} \Phi_2\bigr)^2 + h.c) + \dfrac{m^2_s}{2}\Phi^2_s + \dfrac{\lambda_6}{8}\Phi^4_s + \dfrac{\lambda_7}{2}\Phi^2_s\abs{\Phi_1}^2 + \dfrac{\lambda_8}{2}\Phi^2_s\abs{\Phi_2}^2.
\end{align}
This potential is invariant under a $Z_2$ symmetry which acts only on the real singlet scalar $\Phi_s \rightarrow -\Phi_s$. The scalar fields obtain a vacuum expectation value (VEV)\footnote{Non-zero $v_+$ corresponds to electric charge violating vacua and $\xi \neq 0$ corresponds to CP-violating vacua. We only focus on neutral vacua on the boundaries of the wall ($v_+ = 0$ and $\xi =0$).}:
\begin{align}
   \langle \Phi_1 \rangle = \text{U} \dfrac{1}{\sqrt{2}}
    \begin{pmatrix}
          0 \\      v_1
     \end{pmatrix},      
&& \langle \Phi_2 \rangle  = \text{U} \dfrac{1}{\sqrt{2}}
      \begin{pmatrix}
     v_+ \\
     v_2e^{i\xi}
      \end{pmatrix} , && \langle \Phi_s \rangle = v_s,
&& \text{U} = e^{i\theta} \text{exp}\biggl(i\dfrac{\tilde{g}_i\sigma_i}{2v_{sm}}\biggl),      
\label{eq:vacuumform}      
\end{align}
where U is an element of the $\text{SU(2)}_L\times\text{U(1)}_Y$ with $\theta$ and $\tilde{g}_i$ denoting the Goldstone modes of the scalar doublets, $\sigma_i$ the Pauli matrices and $v_{sm} \approx 246 \text{ GeV} $ the standard model vacuum expectation value (VEV).
The real singlet scalar acquires a non-zero minimum in the early universe which breaks the $Z_2$ symmetry and leads to the formation of a domain wall network that interpolates between minima with positive and negative VEVs. This, in turn, makes the terms $\dfrac{\lambda_7}{2}\Phi^2_s\abs{\Phi_1}^2 + \dfrac{\lambda_8}{2}\Phi^2_s\abs{\Phi_2}^2$ in the effective potential of the 2HDM space-dependent. 
Electroweak symmetry breaking is achieved when the effective mass term of the Higgs doublets $M^2_1(x) = m^2_{11} + (\lambda_{3} + \lambda_4 + \lambda_5)\abs{\Phi_2}^2 + \dfrac{\lambda_7}{2}v^2_s(x)$ is negative\footnote{Same behavior for the effective mass term of the second doublet.}. However, inside and in the vicinity of the domain wall $v_s(0)$ is zero (see Figure \ref{subfig:vs}), and $M^2_1$ receives a large positive contribution as $\lambda_7v^2_s$ vanishes. In such a case, $M^2_1$ can become positive (for $m^2_{11}>0$). The potential of the Higgs doublets is therefore in the symmetric phase as illustrated in Figure \ref{subfig:effmass}.
In order to obtain the vacuum field configuration of the doublet we solve the coupled system of equations of motion of the singlet and doublet scalar fields, taking the boundary conditions of $v_s(-\infty)<0$ and $v_s(+\infty)>0$ and of the VEVs $v_1$ and $v_2$ leading to $v_{ew} = \sqrt{v^2_1 + v^2_2} \approx 246 \text{ GeV}$. In order to generate a region with CP-violating vacua in the vicinity of the wall (see \cite{Battye:2020sxy, Law:2021ing,Sassi:2023cqp} for more details), we chose an initial kink profile for $\theta(x)$ such that $\theta(-\infty) = 0$, $\theta(+\infty) = \pi/2$ and the profile interpolating between both values at approx $x \cdot m_h \approx 7$. Such an initial condition would correspond to two regions of the universe acquiring EW vacua with different Goldstone modes. The solutions\footnote{Solved numerically using the Gradient Flow Method \cite{Battye:2020sxy}.} to the system of differential equations is shown in Figure \ref{subfig:sol}. We find that for the used parameter point (see Table 8 in \cite{Sassi:2024cyb}), the EW vacuum vanishes in a large region around the wall and that a region with CP-violating vacua is generated in the vicinity of the wall where $\theta(x)$ sharply changes. Due to the vanishing of the EW VEV $v_{ew}$ around the wall, the weak sphalerons will be active inside that region while exponentially suppressed outside of it. 

Due to the tension of the domain wall, the requirement that the effective mass becomes positive inside the wall is insufficient to induce electroweak symmetry restoration in the core and vicinity of the wall. In practice, the change in the effective mass terms $M^2_{1,2}$ needs to occur at larger regions in space to make the VEVs of the doublets vanish. We found in \cite{Sassi:2024cyb} that parameter points leading to EWSR in a large region around the wall correspond to negative and large ratios of $\lambda_{7,8}/\lambda_6$. In particular, we found that parameter points of the N2HDM, satisfying all experimental and theoretical constraints \footnote{We impose the theoretical constraints of perturbative unitarity, boundedness from below, and vacuum stability as well as the experimental constraints of EW precision measurements, flavour constraints and collider searches. These parameter points are generated using ScannerS \cite{Muhlleitner:2020wwk}. We also impose the restoration of the $Z_2$ symmetry of the singlet scalar in order to produce the domain walls.} and leading to the restoration of the EW symmetry in a large region around the wall, typically have large values of $v_s$ and smaller CP-even Higgs masses. We also found that parameter points with higher singlet admixture in the SM Higgs state lead to the smallest VEVs inside the wall and to larger regions of EWSR. This correlation can put strong experimental constraints on the possibility of EWSR via domain walls in the N2HDM from current and future collider searches. 

A complete study of baryogenesis using the properties of these domain wall solutions is under investigation.

\subsection*{Acknowledgments}
This work is funded by the Deutsche Forschungsgemeinschaft (DFG) through Germany’s Excellence Strategy – EXC 2121 “Quantum Universe” — 390833306.

\bibliographystyle{JHEP}
\bibliography{references}

\end{document}